# The time distribution of aftershock magnitudes, fault geometry and aftershock prediction

Pathikrit Bhattacharya [1], Kamal [1], Bikas K. Chakrabarti [2]

SUMMARY

We have analyzed, for the first time, the time cumulant of magnitudes of an aftershock sequence since the mainshock. This comes out to be a remarkable straight line whose slope is characteristic of the fault zone. This will provide an useful tool in understanding the temporal distribution of aftershocks after a specific mainshock.

[1]Department of Earth Sciences, Indian Institute of Technology, Roorkee-247 667, Uttarakhand, INDIA.
[2]Theoretical Condensed Matter Research Division and Centre for Applied Mathematics and Computational Science, Saha Institute of Nuclear Physics, 1/AF, Bidhannagar, Kolkata-700 064, INDIA.

E-mail addresses: pathipes@iitr.ernet.in (P. Bhattacharya), kamalfes@iitr.ernet.in (Kamal), bikask.chakrabarti@saha.ac.in (B. K. Chakrabarti).

1 INTRODUCTION

All major earthquakes are followed by aftershocks: events, generally of smaller magnitude, originating from the same rupture zone as the mainshock or about within one to two fault lengths near it (Scholz 2002). Here we present an empirical finding describing the statistics of individual aftershock sequences. This law therefore describes a localized statistical feature of earthquakes. In particular, we show that the cumulative integral of magnitudes of aftershocks with respect to the time elapsed since the occurrence of the mainshock is a straight line whose slope is characteristic of the causative fault zone. We show, with examples from all over the world, that this feature is very robust and may provide an useful tool for understanding and extrapolating the temporal distribution of aftershocks for a given mainshock.

2 DATA COLLECTION AND METHOD

We first collected the aftershock magnitude-time sequences $M(t)$ of eleven major earthquakes from different geographical regions of the world. The earthquakes were selected carefully from all over the globe to ensure that no regional bias was introduced due to the choice of a specific catalog or a specific geological setting. We also intentionally selected some multiple events in the same geological region on a) different fault zones b) the same fault zone at a different time. We then evaluated a cumulative integral $Q(t)$ of the aftershock magnitudes over time. Numerically, we evaluated the integral $Q(t) = \int_0^t M(t')dt'$, the time cumulant of magnitude where $t$ denotes the time since the main shock. Aftershocks of a major event were considered to be events within a given region, geographically defined as boxes or polygons constrained by suitable latitudes and longitudes, and the magnitudes were recorded over a length of time (usually a period of 1000 days or more) over which the region has not yet relaxed to its background seismicity. The integration is done over all event magnitudes above a threshold level, in this case chosen to be the completeness magnitude $M_c$. The datasets we used for our analyses are as follows:

a) The 1989 Loma Prieta earthquake (18/10/1989, $M_w$ = 7.1, 37.0°, - 121.88°). The data set used was the same as the one used in (Kamal & Mansinha 1996).

b) the 1995 Kobe earthquake (17/01/1995, $M_{JMA}$ = 7.2, 34.6°, 135.0°). The aftershock region was chosen on the basis of the work of Toda *et al.* (Toda *et al.* 1998) (latitudes 34°-36°, longitudes 133.5°-137°). The data were taken from the JUNEC catalog for the period 17/01/1995-31/12/1995.

c) The 2004 Sumatra earthquake (26/12/2004, $M_w$ = 9.0, 3.30°, 95.98°). The box was chosen in accordance to the earthquake summary poster prepared by the USGS (available at http://earthquake.usgs.gov/eqcenter/eqarchives/poster). The box chosen was latitudes 0°- 20°, longitudes 90°-100°. The data were taken from the USGS (PDE) catalog for the period 26/12/2004-28/05/2008.

d) The Muzaffarabad (Kashmir, North India) earthquake of 2005 (08/10/2005, $M_S$ = 7.7, 34.52°, 73.58°). The box chosen is defined by latitudes 33.5°-35.5° and longitudes 72.2°-74.2°. The data were once again from the USGS catalog for the period 08/10/2005-28/02/2008.

e) The Chamoli earthquake (29/03/1999, $M_S$ = 6.6, 30.51°, 79.40°), the aftershocks were obtained from a highly localized network employed by the Wadia Institute of Himalayan Geology (Kamal & Chabak 2002).

f) The Bam earthquake (26/12/2003, $M_S$ = 6.8, 29.00°, 58.31°). The box was chosen to be latitudes 27.5°-30.5° and longitudes 57.5°-59.5°, the time interval being from 26/12/2003-26/12/2005 on the IIEES listing reported only in local magnitude $M_L$.

g) The Zarand earthquake (22/02/2005, $M_S$ = 6.5, 30.80°, 56.76°), the aftershock sequence was chosen over latitude extent 29.5°-32.5° and longitude extent 55.5°-59.5° for the time period 22/02/2005-22/02/2007. The catalog used was again IIEES.

h) The Denali fault earthquake in central Alaska (03/11/2003, $M_S$=8.5, 63.52°, -147.44°). The listing was taken from the PDE listing. The box was: latitude 65.0°-60.0° and longitude -141.0°–(-151.0°) and the time window was 23/10/2002-02/05/2008. This sequence is referred to as Alaska 1 henceforth in the text.

i) The Rat Islands, Aleutian Islands earthquake in Alaska (17/11/2003, $M_W$ = 7.8, 51.15°, 178.65°). The source was again the PDE catalog. The latitude longitude box was defined by latitudes 54°-50° and longitudes 174°–(-174°) and the time series was taken between 17/11/2003-02/05/2008. This dataset is henceforth referred to as Alaska 2.

j) The Taiwan earthquake (31/03/2002, $M_W$ = 7.1, 24.13°, 121.19°) for which the listing was taken from BATS (Broadband Array in Taiwan for Seismology) CMT catalog. The relevant box was latitudes 20°-23° and longitudes 119°-122° for the time period 31/03/2002-31/03/2005. This sequence is called Taiwan 1 in the text.

k) Another Taiwan earthquake on the same plate boundary, viz. the Eurasian plate and the Philippines plates, (26/12/2006, $M_W$ = 6.7, 21.89°, 120.56°). The latitude longitude box was defined as 20°-23° and 119°-122° and the time window was 26/12/2006-23/05/2008. The catalog was once again the BATS CMT catalog.

(All dates are expressed as dd/mm/yyyy, positive latitude implies northern hemisphere, positive longitude implies eastern hemisphere, negative latitude implies southern hemisphere and negative longitude implies western hemisphere.) $Q(t)$ for each of the above aftershock sequences were estimated using the above data sets.

The important limitation of our analysis, while evaluating the aforementioned integrals, is the fact that more often than not most catalogs which give the most exhaustive list of aftershocks report the various events in different magnitude scales. This warrants the need for using conversion relationships to convert the various magnitude scales to a uniform scale. This, wherever we have inhomogeneous catalogs, we have chosen to be $M_w$, the moment magnitude as defined by Kanamori (Kanamori 1977). To this end we have used well defined and previously employed conversion relationships. The datasets extracted from the NEIC (PDE) catalog are all inhomogeneous with respect to the magnitude scales used to report the various events. The PDE listing was used in case of the Sumatra, Muzaffarabad and the Alaska events (see Table 1). For the Sumatra event we used the conversion relationships used in (Jiang & Wu 2006). These relationships were specifically designed for the aftershock sequence of the Sumatra event extracted from the PDE catalog and hence serve our purpose. For the Muzaffarabad event we used conversion relations given in (PMD & Norsar report 2007) which were again designed specifically for the region and is based on the PDE listing. The fact that the conversion relationships were designed for nearly the same datasets as we have used here is important as such conversion models are in general regressional models and hence their use in our work is validated by the fact that here we use them on the same population for which they were originally designed. For the two aftershock sequences Alaska 1 and Alaska 2 we could not obtain valid conversion relationships. Our strategy for these two sequences is described later in the text. Once we had homogeneous magnitude-time listings, we calculated the *b*-values and completeness magnitudes ($M_c$) for each of the recorded aftershock sequence. Our estimate of $M_c$ is based on the assumption that the Gutenberg Richter (GR) law (Gutenberg & Richter 1944) explains a large percentage of the frequency-magnitude distribution above a given completeness magnitude. We assumed the percentage to be an ad hoc 90% and followed the methodology in (Wiemer & Wyss 2000). It has been recently noted (Wiemer & Wyss 2002) that the assignment of a single completeness magnitude to an aftershock sequence is oversimplified and the $M_c$ for such sequences have temporal variability. Within the first hours to days of an aftershock sequence, $M_c$ tends to decrease systematically. This is caused by improvements of the station network and the fact that the frequent larger aftershocks eclipse smaller events. In our case however, as we track an aftershock sequence over a large period of time exceeding at least a year, this initial variability has been disregarded and a unique $M_c$ has been ascribed to each of the homogenized sequences. Such an assignment is problematic for the Alaska 1 sequence due to the lack of homogeneity (and hence increased possibility of bias in determination of *b*) discussed previously. So we progressed following a different methodology. For Alaska 1, the most numerous listing of events was in $M_L$. So we chose this subset of the data (1848 events out of a total 2031 events) and calculated the *b*- value and $M_c$ on this set. This was done keeping in mind that the lower magnitudes are more regularly reported in $M_L$ in the PDE catalog. However, for the Alaska 2 listing no such clear maximal homogeneous subset was available and we calculated $M_c$ and *b* based on the entire inhomogeneous dataset. This is followed by calculation of $Q(t)$ where the cumulative integral is carried out only over events that have magnitude $M \geq M_c$. Therefore in our case the minimum magnitude or lower cut off is $M_c$. Our analyses indicate a clear linear relationship $Q(t) = St$ where *S* denotes the slope. Wherever we have a homogeneous listing, we have attempted to fit a straight line to the observed $Q(t)$ curve and we call the slope thus obtained as $S_1$. Wherever we have a inhomogeneous listing the slope obtained by fitting is renamed as $S_2$.

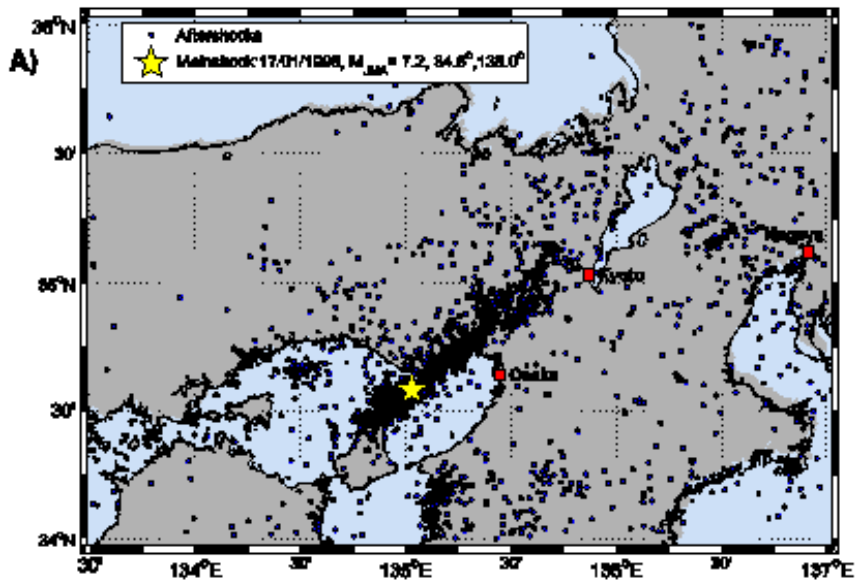
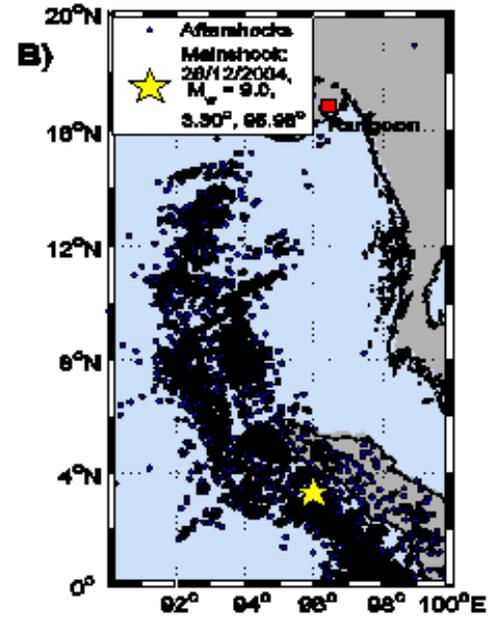
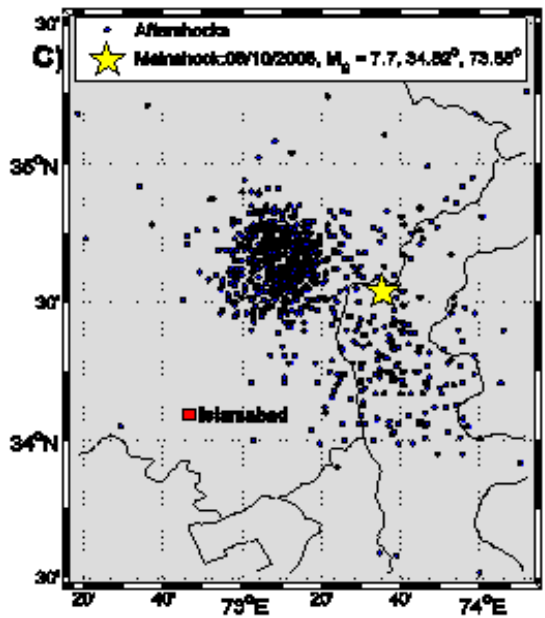
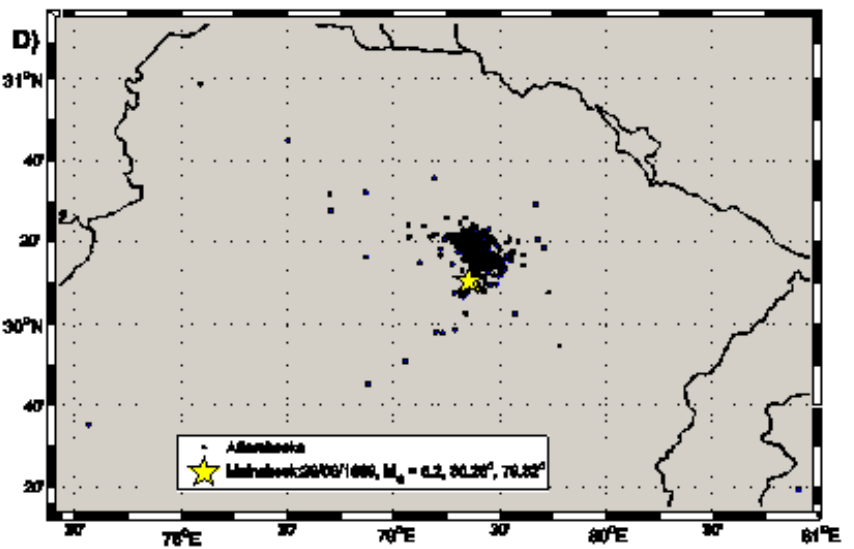
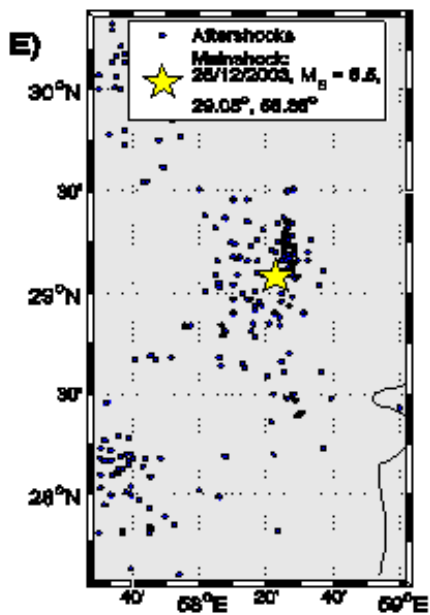
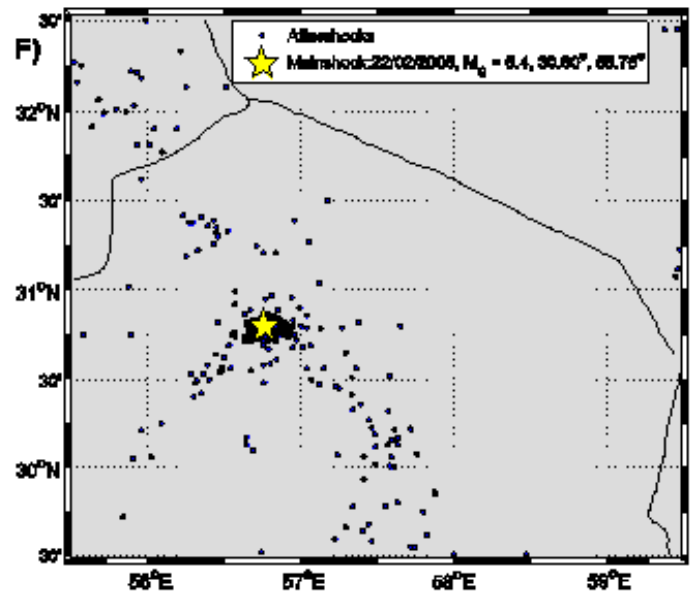

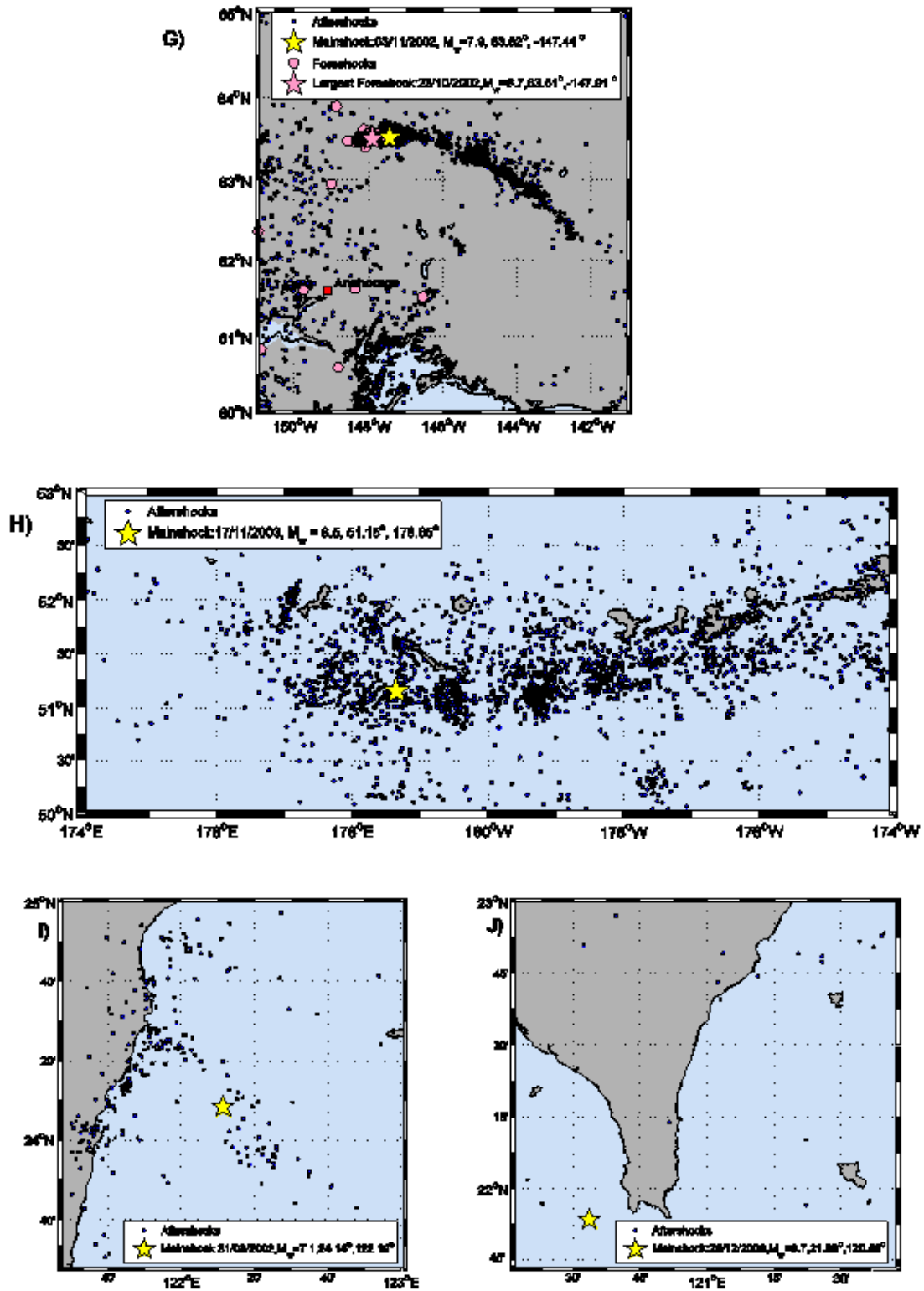

Fig.1. Maps showing spatial distribution of aftershocks chosen for each of the above data sets except for Loma Prieta. The indexing of the maps is according to the listing of the datasets i.e. A) Kobe, B) Sumatra, C) Muzaffarabad, D) Chamoli, E) Bam, F) Zarand, G) Alaska 1, H) Alaska 2, I) Taiwan 1 and J) Taiwan 2.

| Event Name (Event Tag) | $M_c$ | b | $S_1$ | $S_2$ | RMS error in fitting | R square |
|---|---|---|---|---|---|---|
| Loma Prieta (L) | 0.50 | 0.67 | 1.15 | - | 2.22 | 0.99 |
| Kobe (K) | 2.00 | 0.91 | 2.41 | - | 1.11 | 0.99 |
| Sumatra (S) | 3.63 | 0.81 | 4.16 | - | 5.18 | 0.99 |
| Muzaffarabad (M) | 3.33 | 0.79 | 4.05 | - | 7.38 | 0.99 |
| Chamoli (C) | 0.20 | 0.32 | 1.39 | - | 2.02 | 0.99 |
| Bam (B) | 2.70 | 0.90 | 3.48 | - | 45.28 | 0.99 |
| Zarand (Z) | 2.80 | 0.99 | 3.47 | - | 10.11 | 0.99 |
| Alaska 1 (Al 1) | 3.00 | 1.08 | - | 3.47 | 12.78 | 0.99 |
| Alaska 2 (Al 2) | 2.60 | 0.46 | - | 3.47/4.08 | 9.41/11.37 | 0.99/0.99 |
| Taiwan 1 (T 1) | 3.42 | 0.77 | 4.30 | - | 39.55 | 0.99 |
| Taiwan 2 (T 2) | 3.63 | 0.75 | 4.32 | - | 17.73 | 0.99 |

*Table 1. Event names are used to refer to respective sequences in the text. The event tags correspond to those in the plot in Fig.2 and Fig.3. $S_1$ corresponds to the slope of the linear fit with the homogeneous listings while $S_2$ corresponds to the linear fit with the inhomogeneous datasets. In Alaska 2, the slope changes midway (see Fig.4c) and the two slopes, rms errors and R square values depict the values obtained while fitting for the earlier part before the slash and for the later part after the slash.*

This is the nomenclature followed in Table 1 in listing the slopes for each sequence. The results of our analysis mentioned in Table 1 and the plots in Fig. 3 point clearly to the linear variations mentioned above. The straight line (fit) retains this slope for years. Also, the slope changes significantly over different fault zones. This indicates that the slope, $S$, is characteristic of the fault zone. This was further checked by integrating from anywhere in the time series (i.e. shifting our $t = 0$ to any randomly chosen aftershock) after the mainshock. The slope was found to be the same and the linearity of $Q(t)$ was not affected by shifting the origin of integration. A wide variety of events can lead to systematic errors in the reported magnitudes (events as varied as a change in instrumental calibration to addition or removal of seismograph stations) and such systematic errors can be very large going up to as much as 0.5 magnitude units (Castellaro *et al.* 2006). Such errors would set the eventual error bound for the slope as the errors due to fitting are much smaller as mentioned already. Additionally, the conversion relationships themselves induce some errors in the magnitudes. This can also lead to systematic errors in the slope estimate. With the available catalogs, the errors in slope estimation would be thus about 6-10% *(Castellaro et al. 2006, Sipkin et al. 2000).*

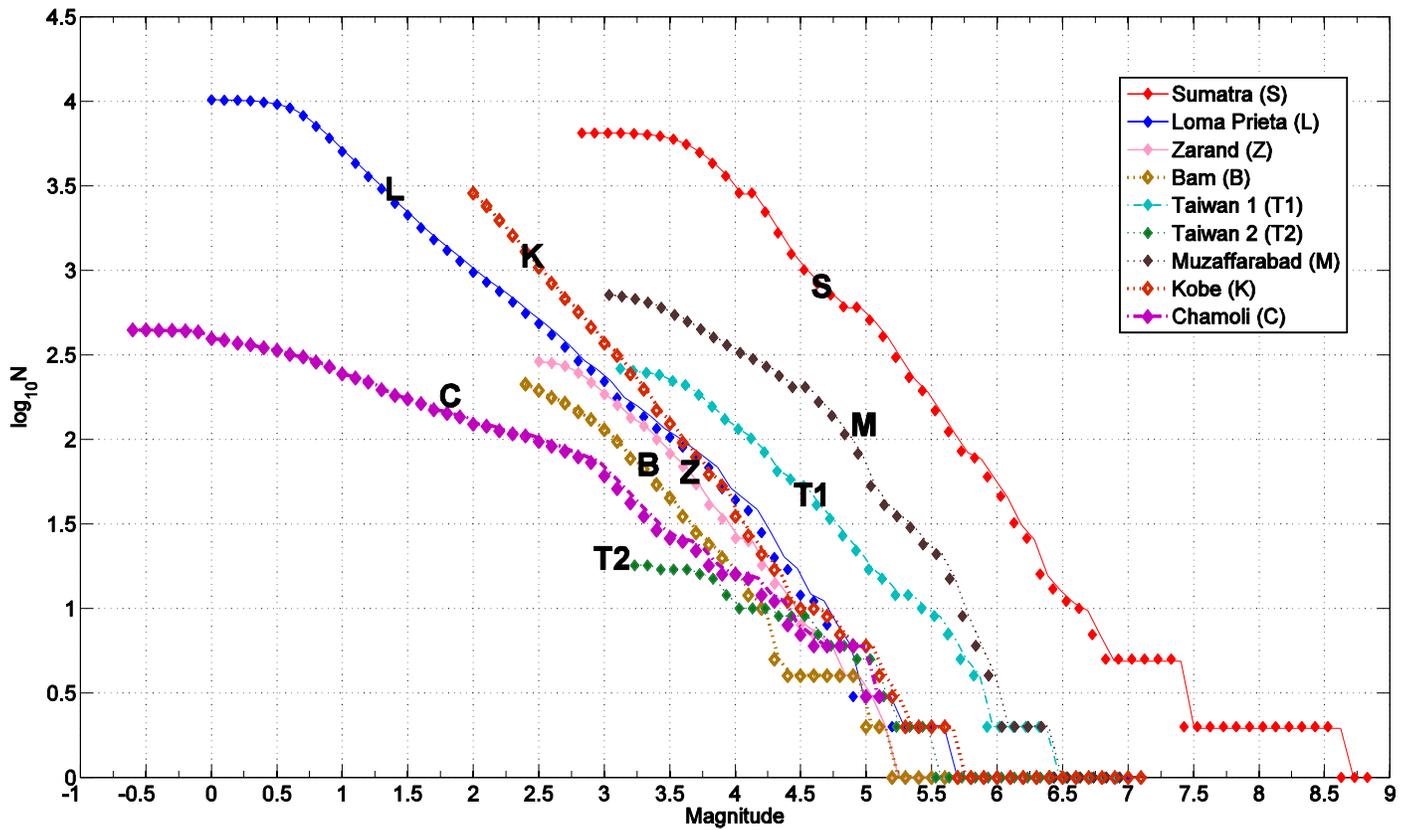

*Fig. 2. The Frequency-Magnitude distributions for the various aftershock sequences enlisted in Table 1 (except for the Alaska 1 and Alaska 2 sequences)*

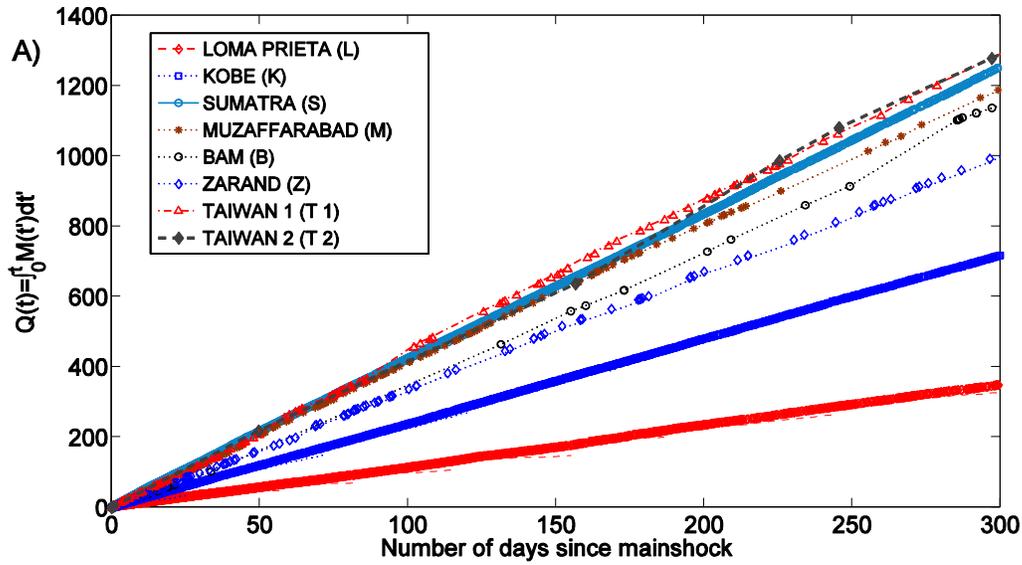

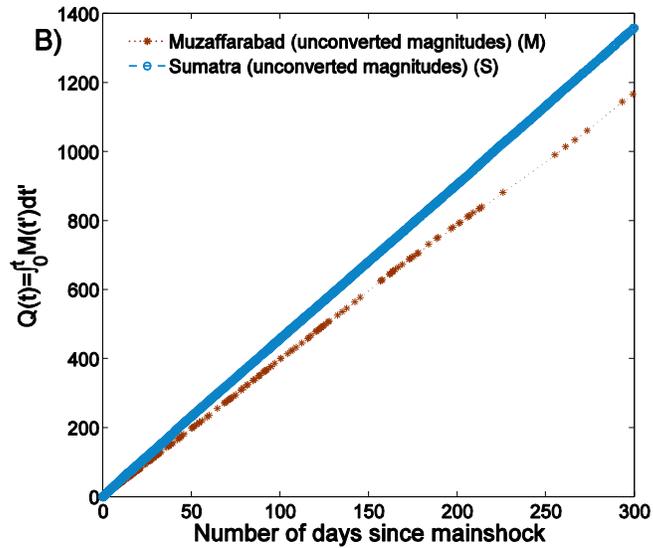

*Fig.3. (A) Plots of time cumulant of magnitude Q(t) vs. t (in number of days since the mainshock) for the datasets described in the text and in Table 1 for the first 300 days. The tags for the events used in the plot are the same as in Table 1. (B) Plots of Q(t) Vs. t for the Sumatra and Muzaffarabad sequences before conversion of magnitudes according to (Jiang & Wu 2006) and (PMD & Norsar report 2007) respectively. Again, only the first 300 days are plotted.*

3 ANALYSIS OF THE RESULTS

As we said earlier, it is our contention that these slopes are characteristic of the fault zone. Firstly, there is significant global variation in the values of the slopes. Each aftershock sequence seems to have a different slope for the $Q(t)$ statistic. Significantly, aftershock sequences caused by the same fault zone or on the same geologic and tectonic setting yield similar slopes. These two facts are clear from the results in Table 1. To illustrate the second point further, we draw attention to the two Taiwan sequences, Taiwan 1 and 2. Both of these took place on the Eurasian and Philippines plate boundary (USGS summary poster for the event). It is only natural that the two corresponding slopes would be nearly identical in view of our proposed error bounds due to the geological similarities and precisely similar tectonism. In Iran though, on the contrary, the Bam and Zarand earthquakes took place on two different faults belonging to a highly developed fault system. The Bam event occurred on the Bam fault whereas the Zarand event took place in close proximity of a previous event on the Gowk system (1981 July 28, Sirch earthquake $M_W = 7.1$) at a distance of about 60 km from the northern extremity of the rupture zone of the Sirch event (Nalbant *et al.* 2006). But still the slopes were found to be the same (within proposed error bounds). This might be because the Gowk system and the Bam fault are part of a highly developed fault system. Further, the slope does not change with unusually large aftershocks in the sequence e.g. the Sumatra sequence had a few very large aftershocks including one great earthquake on March 28, 2005 ($M_W = 8.7$) which occurred about 150 km SE of the earlier giant earthquake epicenter ($M_W = 9.3$) of December 26, 2004. This further reveals the characteristic nature of the slope.

More importantly it is critical to note that the integration process is akin to an averaging process on the magnitudes and the slope, as we will later see, is approximately an average magnitude. So the reader may immediately think that the slope is more a signature of the catalog and the completeness level of the listing rather than the fault itself. But then if we have examples where reactivation or re-rupturing or similar events on a preexisting fault system changes the slope of $Q(t)$ over time, that would establish the characteristic nature of the slope. The reason for this is two fold. Firstly, as we will show later, we have considerable reason to believe that the slope of the $Q(t)$ statistic depends significantly on the geometry of the faults and the slope change is an expected result of the change in geometry of the fault surface due to the re-rupturing. Secondly, the inherent changes in completeness level in aftershock sequences are limited to within days of the occurrence of the mainshock. If we observe a change in slope a long time after the mainshock, when the $M_c$ has stabilized to a steady level, then this change in slope cannot be ascribed to a change in completeness level. This slope change is then due to the changes in asperity distribution and stress patterns brought about by the re-rupturing. If the slope changes due to these factors then of course it is characteristic of the fault involved. This part will become very clear later when we discuss the statistical interpretation of the slope as average magnitude. At present let us look at an example of such an occurrence which strengthens our claims. To this end we shall use the results for the two sequences obtained in Alaska. Alaska 1 was an event on the inland Denali fault and the $Q(t)$ statistic gives a slope $S_2 = 3.47$. One important aspect came out during the analysis of the Alaska 1 dataset. The first shock considered here was not the Denali fault mainshock but a previous shock in the same region. This was done because this event is a very well established foreshock of the Denali fault event and is followed by a series of foreshocks till the occurrence of the mainshock. The cumulant $Q(t)$ for the foreshocks retains the same linearity as the cumulant $Q(t)$ for the aftershocks (see Fig.4b). This is expected though if one believes in the characteristic nature of the slope for a given fault system. As foreshocks and aftershocks both happen on the same fault or fault system they are expected to yield the same slopes. Our claim, that the slope is a signature of the fault system, is further strengthened on analysis of the Alaska 2 aftershock sequence (see Fig. 4c). Here, the slope of the $Q(t)$ vs. $t$ curve increases after about 754 days of the main event. One of the most significant events of the last century, the 1965 $M_w$ 8.7 Rat Islands earthquake ruptured a ~600 km-long portion of the plate boundary to the west of the Amchitka Island. In the November 17, 2003 M7.7 earthquake, the main shock or the first shock in the sequence we chose, the easternmost part of the 1965 zone failed again. On June 14, 2005, a series of moderate to strong earthquakes occurred in the Rat Islands region of the Aleutian Islands. The sequence started with a M5.2 event at 08:03 UTC and the largest event of M6.8 followed 9 hours later (at 17:10 UTC). The largest earthquake was situated 49 kilometers (31 miles) south-southeast of Amchitka. This new sequence of earthquakes re-ruptured the easternmost end of the 1965 rupture zone. This is the reason, we believe, for the increase in slope. The re-rupturing process meant that the earlier asperity distributions were changed and hence the region underwent a marked change in its seismicity pattern. The slope measures for the Alaska 2 sequence are given in Table 1 for the total inhomogeneous event listing i.e.

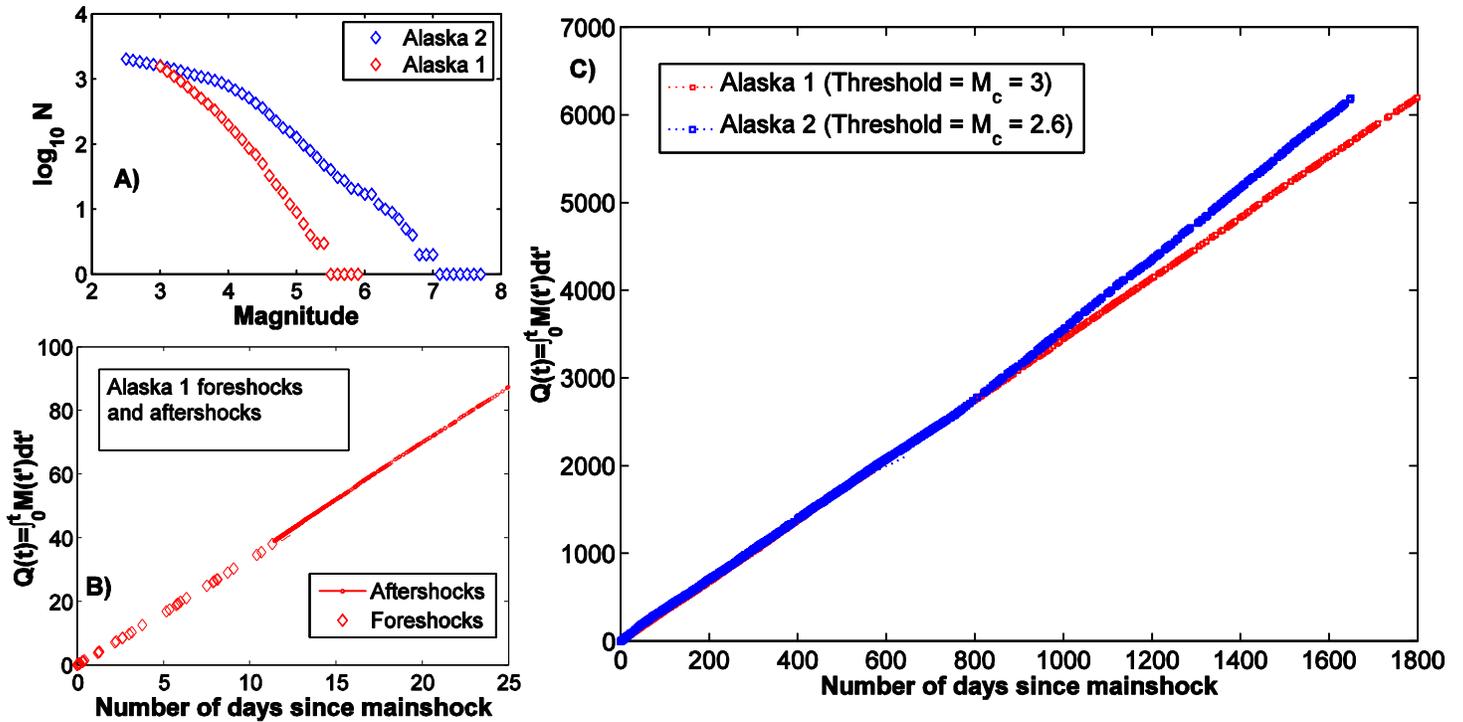

*Fig.4. A) The frequency magnitude distribution (FMD) plots for Alaska 1 and Alaska 2. As noted in the text, the FMD for Alaska 1 is computed on the $M_L$ listing. For Alaska 2 the FMD was calculated on the entire inhomogeneous listing. B) Q(t) for the first 25 days for Alaska 1 showing that foreshocks and aftershocks exhibit same slope. C) The entire sequences for Alaska 1 and 2. Note slope change in Alaska 2.*

$S_2 = 3.47$ for the earlier half and $S_2 = 4.08$ for the later half. The change in slope, as one can clearly observe is not within our error bounds. The slope change also occurs after about two years from the mainshock and hence should not be an artifact of change in minimum completeness level of the seismic network which is constant by now. Also the slope does not keep on changing and remains at $S_2 = 4.08$ for the rest of the sequence. The slope of $Q(t)$ definitely shows that some significant change occurred in the aftershock sequence (equivalently the fault system) due to re-rupturing.

However such cumulative statistics have already been attempted for the scalar seismic moment or Benioff stresses for aftershock sequences. We did a similar cumulative integral of scalar seismic moment for our sequences in Sumatra and Taiwan (the former was reliably converted to scalar seismic moment in (Jiang & Wu 2006) and the BATS CMT catalog for Taiwan was homogeneous and listed only broadband $M_W$ values). The results are shown in the Fig.5. The resultant plots resemble a step function. Authors (Jiang & Wu 2006) have tried to fit a power law and/or linear models piecewise to such data (in their case the cumulative Benioff stress). There seems to be no robust feature to this statistic, i.e. the cumulative moment versus time curve. Such cumulative curves have also been reported for theoretical models such as for the Critical Continuum-State Branching Model of Earthquake Rupture (Kagan 2006). Precursory accelerating moment release before large earthquakes has been a widely discussed phenomenon until recent years, being regarded as observational evidence for the controversial critical-point-like model of earthquake generation (Sornette & Sammis 1995, Jaumé & Sykes 1999 ). Another useful property of such seismic moment cumulants is that they help in monitoring the stress release modes for a given region and hence allow for discussions on the type of mechanisms underlying earthquake occurrences (Fukuyama *et al.* 2001).

4 INTERPRETATION OF THE SLOPE

As mentioned earlier, the slope of $Q(t)$ is an estimate of the average magnitude of the aftershock sequence. We proceed on this line and try to obtain an expression connecting the slope and other parameters of aftershock statistics. For an aftershock sequence, apart from the first few aftershocks, let us assume that the event inter-occurrence times and their magnitudes are statistically independent of each other in the long term. Then for a large number of events we have,

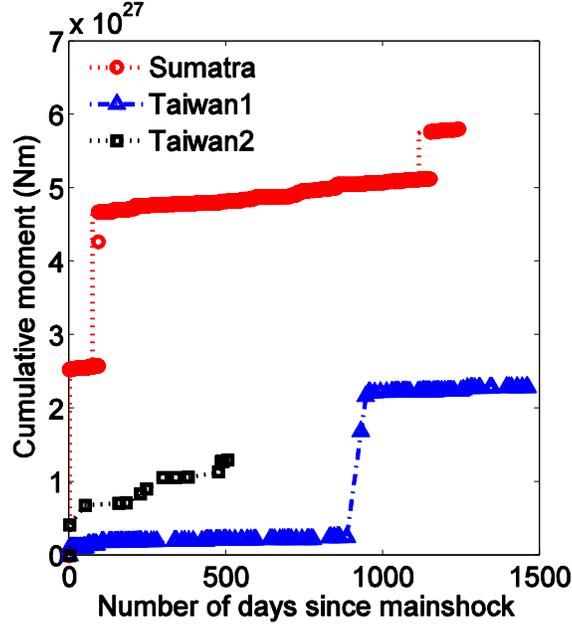

*Fig.5. Plots of cumulative moment versus time since the mainshock for the datasets Sumatra, Taiwan 1 and Taiwan 2. The values for Taiwan 2 depicted here in the plot are 10 times the real values to ensure proper legibility of the figure.*

$$Q(t) = \int_0^t M(t')dt' \approx \int_0^t M_{avg}(t')dt' \approx M_{avg} t \qquad (1)$$

where $M_{avg}$ is the average magnitude calculated from the GR distribution for the aftershock sequence (assuming that the $b$ value is constant over time which is an observed fact (Wiemer & Wyss 2002)). Now this $M_{avg}$ can be calculated from the GR distribution as follows,

$$M_{avg} = \int_{M_{min}}^{M_{max}} M \frac{b \times \ln 10 \times 10^{-bM}}{(10^{-bM_{min}} - 10^{-bM_{max}})} dM \ . \qquad (2)$$

If we assume that the GR law holds for all larger magnitudes then $M_{max}$ tends to infinity and we have,

$$S = M_{avg} = M_{min} + \frac{0.43}{b} \ . \qquad (3)$$

This is the expression for the slope which connects it with the b value for the aftershock sequence. We have presented a comparison of the slopes obtained through fitting and $M_{avg}$ in Table 2 (except for the Alaska 2 sequence because of the complete inhomogeneity and change in slope midway). The results show a very good agreement. But the agreement is in general better for b as computed for the specific aftershock sequence rather than the global b value of unity. In some cases like in Loma Prieta, Taiwan 2 and Chamoli the error (i.e. $S - M_{avg}$) for $b \neq 1$ is an order of magnitude less than the errors obtained when $M_{avg}$ is computed using $b = 1$. Clearly, in general $S = M_{avg}(b)$ is a more accurate expression than $S = M_{avg}(b = 1)$. This brings us to a very important realization. As can be seen in Table 1 and is also reported widely (Wiemer & Wyss 2002), the b values for individual aftershock sequences are unique to the sequence in general and show wide variations. Moreover the b value for a single aftershock sequence has been known to vary spatially between the extremities of the rupture (Wiemer & Wyss 2002). However, temporally, the b value is exceptionally robust except at times when we have rather large earthquakes. Now it is known that the b value of a given aftershock sequence can be explained in terms of the surface similarity dimension of the causative fault network (the fault network being treated as a fractal) (Turcotte 1997). More recently a class of models, referred to as fractal overlap models, give a totally new meaning to the *b* value. The two main models in this class, Self Affine Asperity model (De Rubeis *et al.* 1996) and the Two Fractal Overlap model (Chakrabarti & Stinchcombe 1999, Bhattacharya 2006, Bhattacharya *et al.* 2009), treat relative fault motion as two fractal surfaces sliding over each other and the sticking and slipping of these fractal surfaces cause release of energy as earthquakes. Both these models clearly demonstrate the GR law for the synthetic aftershock sequences and show that the *b* value is purely dependent on the dimension and generation of the fractals involved in sticking and slipping. In the case of real aftershock sequences this would be equivalent to saying that the *b* value is dependent on the fractal geometry of (a) the fault surfaces (b) the fault network. There is overwhelming theoretical and experimental evidence for this. Therefore, the *b* value would change temporally after a large earthquake due to re-rupturing or extensive change in asperity distributions. Re-rupturing would affect the fractal geometry of the fault and in fact in (Bhattacharya *et al.* 2009) it has been shown theoretically that this can only increase the slope of the $Q(t)$ statistic (not decrease). This is what we believe happened in Alaska 2. So as *S* is a function of the *b* value it is characteristic of the fault geometry. Also one must keep in mind that our definition of the completeness magnitude is dependent on the roll-off of the frequency-magnitude distribution from the GR law curve towards the lower magnitude range. Our definition of a roll-off or a fitting algorithm (like the one we used), based on creating synthetic distributions with the *b* value obtained, depend on the value of *b* and a slightly smaller or larger *b* can alter our $M_c$ calculations and hence affect our chosen $M_{\min}$. Moreover, a look at Table 1 also reveals that the slope can be different in spite of similar $M_c$ and vice-versa. Thus the slope can be clearly called a definite characteristic of the fault zone as it is highly affected by the *b* value. The temporal stability of the *b* value explains the constancy in slope over long periods of time.

| Event Tag | $M_c$ | b | $M_{avg}(b=1)$ | $M_{avg}(b)$ | S | S- $M_{avg}(b=1)$ | S- $M_{avg}(b)$ |
|---|---|---|---|---|---|---|---|
| L | 0.50 | 0.67 | 0.93 | 1.14 | 1.15 | 0.22 | 0.01 |
| K | 2.00 | 0.91 | 2.43 | 2.47 | 2.41 | -0.02 | -0.06 |
| S | 3.63 | 0.81 | 4.08 | 4.18 | 4.16 | 0.08 | -0.02 |
| M | 3.33 | 0.79 | 3.78 | 3.90 | 4.05 | 0.27 | 0.15 |
| C | -0.20 | 0.32 | 0.33 | 1.23 | 1.39 | 1.06 | 0.16 |
| B | 2.70 | 0.90 | 3.13 | 3.18 | 3.48 | 0.35 | 0.30 |
| Z | 2.80 | 0.99 | 3.23 | 3.24 | 3.47 | 0.24 | 0.23 |
| Al 1 | 3.00 | 1.08 | 3.43 | 3.40 | 3.47 | 0.04 | 0.07 |
| T 1 | 3.42 | 0.77 | 3.95 | 4.03 | 4.30 | 0.35 | 0.27 |
| T 2 | 3.63 | 0.75 | 4.16 | 4.30 | 4.32 | 0.16 | 0.02 |

*Table 2. The comparison of S and $M_{avg}$ computed from (3) using b = 1 (values listed as $M_{avg}(b=1)$) as well as using b as computed by us from the sequences (values listed as $M_{avg}(b)$). $M_{min} = M_c$ for all the sequences. The errors are listed under S – $M_{avg}$.*

The other question that we tried to address is: Is equation (3) an accurate expression for *S* or a good approximation? To check this we used the following observation. If we increase our $M_{\min}$ by some $\Delta M_{\min}$ above the minimum level of completeness then, we have, using (3),

$$\Delta S = \Delta M_{\text{avg}} = \Delta M_{\min} + \Delta\left(\frac{0.43}{b}\right). \tag{4}$$

Now as $b$ is constant above the completeness level the second term on the right hand side goes to zero and we have,

$$\Delta S = \Delta M_{\min}. \tag{5}$$

We checked for this by recalculating $Q(t)$ for three different minimum magnitude thresholds for each sequence. We chose $M_{\min} = M_c + \Delta$, $\Delta = 0, 0.1$ and $0.2$ as thresholds for each sequence. Nowhere did the results agree exactly to (5). Close agreement was seen only for Alaska1. The results are reported in Table 3. This implies that (3) might be an over-simplified expression for the slope and is not accurate. But one thing is for sure, (3) is a close approximation for the slope. It explains the dependence of the slope on $b$ very well whereas the dependence of the slope on $M_{\min}$ seems not to be straightforwardly linear, as apparent from (3). Hence the slope is probably not simply a manifestation of the lower magnitude cut-off. Moreover, the failure of (3) and (5) to address the slope increment may be an indication that aftershock magnitudes and inter-occurrence times may not be uncorrelated even long after the mainshock thus invalidating (3).

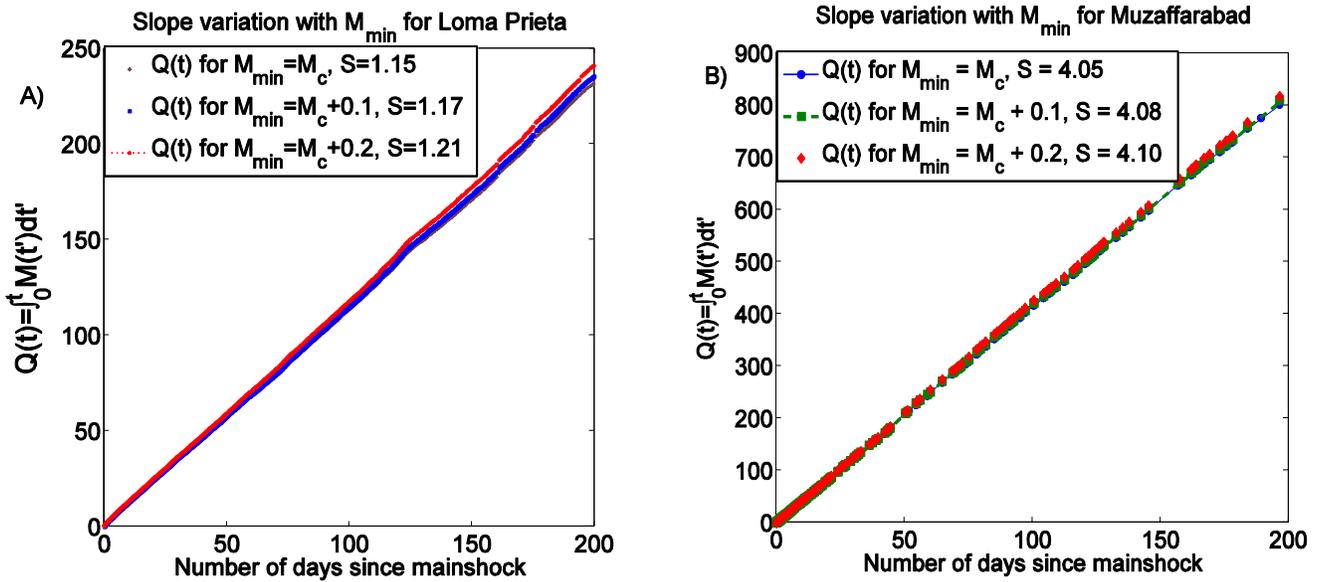

*Fig. 6. Slope variation in Q(t) with $M_{min}$ for the aftershock sequences in Loma Prieta and Muzaffarabad. First 200 days shown for the sake of clarity.*

## 5 NATURE OF THE LINEARITY

To end our discussion we check the nature of the linearity of the $Q(t)$ statistic: Is it deterministic or stochastic? We check this using the following expression for the numerical evaluation of the integral for $Q(t)$;

$$Q(t_i) = \int_0^{t_i} M(t')dt' \approx \sum_i M(t_i)\Delta t_i = Q(t_{i-1}) + M(t_i)(t_i - t_{i-1}) \quad (6)$$

| Event Tag | $M_c$ | S (for $M_{min} = M_c$) | S (for $M_{min} = M_c+0.1$) | S (for $M_{min} = M_c+0.2$) |
|---|---|---|---|---|
| L | 0.50 | 1.15 | 1.17 | 1.21 |
| K | 2.00 | 2.41 | 2.49 | 2.57 |
| S | 3.63 | 4.16 | 4.21 | 4.28 |
| M | 3.33 | 4.05 | 4.08 | 4.10 |
| C | -0.20 | 1.39 | 1.40 | 1.47 |
| B | 2.70 | 3.48 | 3.53 | 3.59 |
| Z | 2.80 | 3.47 | 3.48 | 3.55 |
| Al 1 | 3.00 | 3.47 | 3.56 | 3.63 |
| T 1 | 3.42 | 4.30 | 4.31 | 4.38 |
| T 2 | 3.63 | 4.32 | 4.42 | 4.43 |

*Table 3. Effect of the increment in $M_{min}$ on the fitted slope of Q(t) for all the sequences in Table 2.*

which leads to the expression

$$t_{i+1} = \frac{M(t_{i+1})t_i - Q(t_i)}{M(t_{i+1}) - S} \quad (7)$$

where we make use of the fact that $Q(t_{i+1}) = St_{i+1}$. For all the aftershock sequences we have with us $M(t_{i+1})$ as well as $t_{i+1}$. So starting with the mainshock and using the magnitudes of every subsequent aftershock we could make a synthetic occurrence time listing of using (7) for each $i$ and $i+1$. We calculated the quantity $t_{pred} - t_{ac}$ and created scatter plots against $t_{ac}$. Here $t_{pred}$ are our computed occurrence times and $t_{ac}$ are the actual aftershock occurrence times given in the catalog (i.e. the time since the mainshock). The plots had very significant concentration of points around the line $y = t_{pred} - t_{ac} = 0$ but there were also some very large outliers. In general though the number of deviations were quite smaller. The concentration of most of the points around $t_{pred} - t_{ac} = 0$ clearly demonstrates the deterministic nature of the linearity. The details of the behavior of such plots will be discussed by us elsewhere. At present we give two such plots for the Kobe and Muzaffarabad sequences to strengthen our point and to show clearly the extent of the remarkable concentration of points around $y = t_{pred} - t_{ac} = 0$. As pointed out earlier, the linearity in $Q(t)$ is maintained with the same slope if we shift our origin of integration to any aftershock in the sequence. As cumulative integrals are always subject to accumulation of errors, we also recalculated $t_{pred}$ within various time windows within the series where the origin of the window is far from the mainshock. We observed lesser scatter within the time window than that observed within the same time window when the origin of integration was the mainshock indicating lesser error accumulation as expected. The high point of this result is that if we have apriori knowledge of (or we assume) the magnitude of the next aftershock in a sequence then (7) can be used in real time to predict time of occurrence by extrapolation (which makes perfect sense for this linear curve). Conversely we can have prediction tables where we can obtain a list of occurrence times for the next aftershock corresponding to a list of possible magnitudes of the event. We will discuss the prediction issue in more detail elsewhere. In real time calculations we can minimize errors by shifting our origin of integration or in other words putting $Q(t) = 0$ after every few aftershocks and recalculating all parameters in (7) again by again allowing $Q(t)$ to accumulate over the next few aftershocks and so on. This aspect of our study shows that the linearity of the $Q(t)$ statistic is deterministic and not stochastic. This affords it the unique ability to predict occurrence times of aftershocks given an estimate of the magnitude of the next event (which, as far as we know, is not possible by our current state of knowledge).

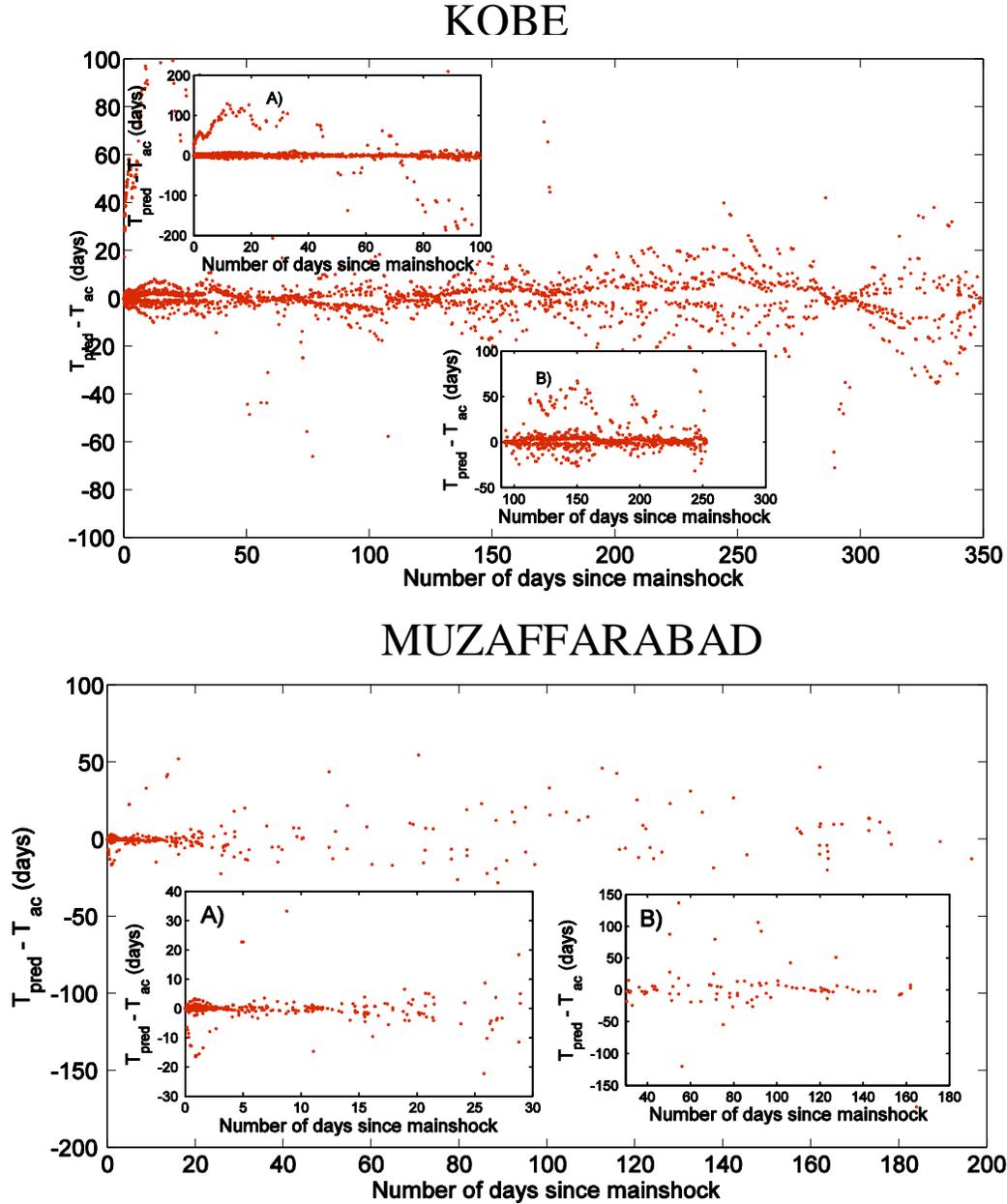

*Fig. 7. The $t_{pred} - t_{ac}$ Vs. $t_{ac}$ plots for the Kobe (top) and Muzaffarabad (bottom) sequences for first 350 and 200 days respectively. Some of the larger deviations from y = $t_{pred} - t_{ac}$ = 0 lie beyond the extent of the y-axis shown here and are omitted to facilitate better viewing. Inset: different time windows within which Q(t) was recalculated. For Kobe: Inset A) First 100 days only. Inset B) Between 90 and 250 days. Scatter lessened significantly with respect to the same time window when we integrate Q(t) from the aftershock at the time origin of the window rather than the mainshock. For Muzaffarabad: Inset A) First 30 days. Inset B) Between 30 to 160 days. Again note the reduced scatter with respect to the same time window when we integrate Q(t) from the aftershock at the time origin of the window rather than the mainshock.*

## 6 CONCLUSIONS

We have shown that cumulative integral of magnitudes $Q(t)$ of an aftershock sequence over time $t$ is linear, the slope $S$ being characteristic of the fault zone. Hence some key features of the rupture zone may be extracted from such an analysis of the magnitude time series $M(t)$ of the earthquake aftershocks, in particular information about the fractal geometry of the fault. We have demonstrated the characteristic property of the slope and its relationship with the $b$ value. We have also shown that though the slope depends on the lower magnitude cutoff, the relationship is not very simple or linear. More importantly we have demonstrated that the linearity of $Q(t)$ is deterministic and hence might very well provide methods to predict aftershock occurrence times in the future.

ACKNOWLEDGEMENTS:

Kamal is thankful to Ministry of Human Resource and Development, India and IITR for extending a kind grant (Grant No.: MHR03-12-801-106 (Scheme B)) which made this work possible.

We are profusely thankful to Lalu Mansinha, Dept. of Earth Sciences, University of Western Ontario for many fruitful discussions and suggestions which helped improve the quality of this manuscript.

Data for the Taiwan events was obtained from Broadband Array in Taiwan for Seismology Centroid Moment Tensor (BATS CMT) catalog available at their website [http://db1.sinica.edu.tw/ ~textdb/ bats/ index.php](http://db1.sinica.edu.tw/ ~textdb/ bats/ index.php).

Earthquake Engineering and Seismology (IIEES), Iran available at their website [http://www.iiees.ac.ir/ EQSearch/(zntqla55rmrcz0jp2na1sa45)/EventQuery.aspx](http://www.iiees.ac.ir/ EQSearch/(zntqla55rmrcz0jp2na1sa45)/EventQuery.aspx).

The catalog for Bam and Zarand was obtained from the International Institute of Earthquake Engineering and Seismology (IIEES), Iran available at their website [http://www.iiees.ac.ir/EQSearch/ (zntqla55rmrcz0jp2na1sa45)/EventQuery.aspx](http://www.iiees.ac.ir/EQSearch/ (zntqla55rmrcz0jp2na1sa45)/EventQuery.aspx).

The JUNEC catalog is available at [http://wwweic.eri.u-tokyo.ac.jp/ CATALOG/ junec/ monthly.html](http://wwweic.eri.u-tokyo.ac.jp/ CATALOG/ junec/ monthly.html).

The NEIC(PDE) (National Earthquake Information Center (Preliminary Determination of Epicenters)) catalog is available at [http://neic.usgs.gov/neis/epic/](http://neic.usgs.gov/neis/epic/).